\documentclass[showpacs,aps,prd,nofootinbib,floatfix,amsmath,amssymb]{revtex4}
\usepackage{graphicx}
\usepackage{multirow}
\usepackage{amssymb}
\usepackage[usenames]{color}
\usepackage[toc,page]{appendix}
\usepackage{slashed}

\newcommand{\decay}{$\tau^- \to \eta^{(\prime)} \pi^- \nu_{\tau}$ }

\DeclareMathAlphabet{\mathpzc}{OT1}{pzc}{m}{it}
\makeatletter
\newbox\slashbox \setbox\slashbox=\hbox{$/$}
\newbox\Slashbox \setbox\Slashbox=\hbox{\large$/$}
\def\pFMslash#1{\setbox\@tempboxa=\hbox{$#1$}
  \@tempdima=0.5\wd\slashbox \advance\@tempdima 0.5\wd\@tempboxa
  \copy\slashbox \kern-\@tempdima \box\@tempboxa}
\def\pFMSlash#1{\setbox\@tempboxa=\hbox{$#1$}
  \@tempdima=0.5\wd\Slashbox \advance\@tempdima 0.5\wd\@tempboxa
  \copy\Slashbox \kern-\@tempdima \box\@tempboxa}

\def\miss#1{\ifmmode{/\mkern-11mu #1}\else{${/\mkern-11mu #1}$}\fi}
\makeatother

\begin{document}

\title{G-parity breaking in \decay decays induced by the $\eta^{(\prime)}\gamma\gamma$ form factor }
\author{G. Hern\'andez-Tom\'e } \email{ghernandez@fis.cinvestav.mx}
\author{G. L\'opez Castro} \email{glopez@fis.cinvestav.mx}
\author{P. Roig}\email{proig@fis.cinvestav.mx}
 \affiliation{Departamento de F\'isica, Centro de Investigaci\'on y de Estudios Avanzados del Instituto Polit\'ecnico Nacional,
Apdo. Postal 14-740, 07000 M\'exico D.F., M\'exico}

\begin{abstract}
Breaking of G-parity or new weak (second class) currents can be responsible for \decay decays. Forthcoming measurements of $\tau$ lepton properties at the Belle II experiment will be able to measure this decay 
channel for the first time. Isolating new physics contributions from the measured rates will require a careful evaluation of G-parity breaking contributions. Here we evaluate the one-loop contribution to \decay decays induced by the emission of two virtual photons 
and its later conversion into an $\eta^{(\prime)}$ meson. As expected, this contribution is very small and may be relevant only for new physics searches contributing at the $10^{-4}$ level to the decay rate of \decay.
\end{abstract}

 \date{\today}
\maketitle
\label{Intro}

\maketitle

\begin{figure}
\begin{center}
    \includegraphics[scale=0.8]{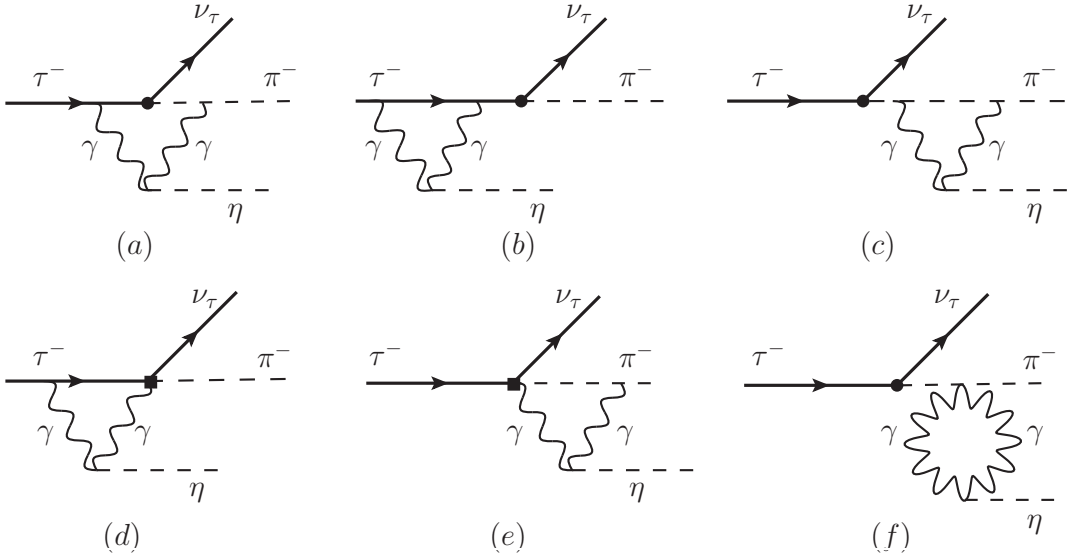}
    \caption{$\tau^- \to \eta^{(\prime)} \pi^- \nu_{\tau}$ decay at one loop level induced by two-photon electromagnetic interaction. The contributions from diagrams (c), (e) and (f) are identically zero (see text). The effective weak $\tau^-\to\nu_\tau\pi^-$ vertex 
    is depicted by a thick dot, while the effective electroweak $\tau^-\to\nu_\tau\pi^-\gamma$ vertex is represented by a thick square.
     \label{dia}}
    \end{center} 
\end{figure}

It is well known that 
electromagnetic and weak hadronic currents with isospin 0, 1 quantum numbers can be classified according to their G-parity~\cite{Lee:1956sw} transformation properties 
into two classes \cite{Weinberg:1958ut}. The first class includes currents 
with quantum numbers $J^{PG}=0^{++}, 0^{--}, 1^{+-}, 1^{-+}$, whereas the second class currents (SCC) have opposite $G$-parity $J^{PG}=0^{+-}, 0^{-+}, 1^{++}, 1^{--}$. 
Since G-parity invariance is broken by isospin non-conservation, electromagnetic effects and the mass difference of $u-d$ quarks can induce the hadronization of the {standard model} (SM) currents into states that mimic the effects 
of  SCC;  therefore, $G$-parity violating processes are 
naturally suppressed. So far, no experimental evidence of SCC weak interactions has been reported. Similarly, isosinglet and isotriplet meson states have well defined G-parity quantum numbers; 
in this case isospin breaking can mix the neutral components of states with 
different G-parity giving rise to the well known $\pi^0-\eta-\eta'$ and $\omega-\rho^0$ mixing phenomena explaining the observed rates of $\omega\to \pi^+\pi^-$ or $\rho\to 3\pi$ decays \cite{Olive:2016xmw}.

A clean test for the existence of SCC would be provided by the observation 
of the semileptonic transitions $\tau^-\to\eta^{(\prime)}\pi^-\nu_\tau$ \cite{Leroy:1977pq}, since the G-parity of the hadronic system ($-1$) is opposite to the one of the charged weak  current in the SM (G=$+1$). Currently, the 
most stringent bounds available are based on searches by the BaBar collaboration \cite{delAmoSanchez:2010pc} corresponding to $BR(\tau^-\to\eta\pi^-\nu_\tau )<9.9\times 10^{-5}$ 
and  $BR(\tau^-\to\eta^{\prime}\pi^-\nu_\tau )<7.2\times 10^{-6}$ \cite{Aubert:2008nj}, which lie close to the estimates based on isospin symmetry breaking \cite{iso} for the 
$BR$(\decay) decays mainly induced by the $u-d$ quark mass difference \cite{ChPT, ChPT2}. Further, Belle II is expected to accumulate up to two orders of magnitude more $\tau$ lepton pairs than BaBar and 
Belle, which should make possible the discovery of SCC.

In addition to the $u-d$ quark mass difference, electromagnetic interactions also break isospin symmetry and will contribute to \decay decays. This can occur at the one-loop level, {\it via} the emission of a pair of photons from $\tau^- \to \pi^-\nu_{\tau}$ decays 
and their later conversion into an $\eta^{(\prime)}$ meson through the anomalous vertex as shown in Figure \ref{dia}
\footnote{Obviously, this type of vertex would also contribute to the $\tau^-\to\pi^-\pi^0\nu_\tau$ decays. However, as we will check, it is 
negligible in this case because the tree level contribution is not suppressed.}. 
Despite this kind of processes are expected to give only a minor correction to the observables, 
it is very important to have a reliable estimate of these effects in order to eliminate a possible source of background for a genuine SCC (see Ref.~\cite{Guevara:2016trs} for 
a dedicated study of the backgrounds given by radiative decays). As a reference to quantify the effect of the new contribution we are studying to the \decay decays, we will 
use the results in Ref.~\cite{Escribano:2016ntp}, which employs a data-driven approach to the vector form factor contributions \cite{Fujikawa:2008ma} and the state-of-the-art 
analysis of meson-meson scattering within unitarized Chiral Perturbation Theory \cite{GuoOller} to obtain those of scalar form factors (see appendix A for definition of those tree-level form factors).

In order to analyze the different contributions for \decay decays induced at one loop level~\footnote{We recall the main formulas for the analyses of these decays at tree level in appendix \ref{A1}.} 
by the electromagnetic interaction we consider, in the low-energy limit, a point-like 
interaction for the $\tau^-\to\nu_\tau\pi^-$ vertex, which can be described by the Lagrangian density

\begin{eqnarray}
\mathcal{L}&=&G_{F}V_{ud}f_{\pi}\bar{\nu}_{\ell}\gamma_{\mu}\left(1-\gamma_{5}\right)\ell\partial^{\mu}\pi^{+}+\textrm{h.c.}\,,
\end{eqnarray}
where $f_{\pi}=F_{\pi}/\sqrt{2}=92.2 \ $ MeV. Form factors complying with the low- and high-energy limits of QCD are considered for the pion electromagnetic coupling and the two-photon coupling of the neutral meson, as it is 
discussed in appendix \ref{A2}, where we also explain the approximations 
adopted in the computation 
of the corresponding loop integrals.

As we mentioned before, we are interested in the study of the $\tau^-(p_{\tau}) \to \eta^{(\prime)}(p_{\eta})\pi^-(p_{\pi})\nu_{\tau}(p_{\nu})$ decays induced at one loop level (see Fig. \ref{dia}) as a possible background for a genuine SCC. 
The decay amplitudes for diagrams Fig. \ref{dia} (c) and (f) vanish owing to the conservation of P and CP by strong and electromagnetic interactions, whereas the contribution from diagram 
(e) vanishes when considering the loop integration because it is odd in the integration variable. After performing the loop integration for diagrams in Fig. \ref{dia} (a), (b) and (d) 
and employing the Chisholm identity \footnote{Chisholm identity reads $\gamma_{\mu}\epsilon^{\mu\nu\rho\sigma}=i\left(\gamma^{\nu}\gamma^{\rho}\gamma^{\sigma}-g^{\nu\rho}\gamma^{\sigma}-g^{\rho\sigma}\gamma^{\nu}+g^{\nu\sigma}\gamma^{\rho}\right)\gamma^5$.} for the 
Levi-Civita tensor contracted with a gamma matrix, we get the following generic form for each of the non-vanishing amplitudes

\begin{equation}
\mathcal{M}^k =\frac{e^4G_FV_{ud}f_{\pi}}{16 \pi^2}g_{\gamma\gamma\eta^{(\prime)}}\bar{u}\left(p_{\nu}\right)\left[F^k_{0}P_{R}+F^k_{1}\slashed p_{\pi}P_{L}\right]u\left(p_{\tau}\right),\label{amplitud}
\end{equation}
where the superindex $k=a,b,d$ labels the non-vanishing Feynman diagrams in Figure \ref{dia}. 

The form factors $F^k_{0,1}$ are generated by the loop integration and will be discussed in more detail below eq.~(\ref{M2}). The factor 
$g_{\gamma\gamma\eta^{(\prime)}}$ corresponds to the value of the $\gamma^{(*)}\gamma^{(*)}\eta^{(\prime)}$ form factor 
for on-shell photons, which is a global dependence of the matrix element (\ref{amplitud}). Its analogue for the $\pi^0$ case, $g_{\gamma\gamma\pi^0}$, is fixed by the ABJ anomaly \cite{ABJ}

\begin{equation}
 g_{\gamma\gamma\pi^0}\,=\,\frac{N_C}{12\pi^2 f_\pi},
\end{equation}
with $N_C=3$ in QCD. In terms of this $\pi^0\gamma\gamma$ coupling, $g_{\gamma\gamma\eta^{(\prime)}}$ are determined  \cite{Roig:2014uja} considering Chiral Perturbation Theory in the large-$N_C$ 
limit \cite{LargeN, ChPTLargeN}, namely

\begin{equation}\label{TFF}
 g_{\gamma\gamma\eta}\,=\,\left(\frac{5}{3}C_q-\frac{\sqrt{2}}{3}C_s\right)g_{\gamma\gamma\pi^0}\,,\quad \ \ \ g_{\gamma\gamma\eta'}\,=\,g_{\gamma\gamma\eta}\left(C_q\to C_{q'},\,C_s\to-C_{s'}\right) \,,
\end{equation}
where the input values for the mixing coefficients can be found in Ref.~\cite{Roig:2014uja}.

The square of the total decay amplitude (${\cal M}=\sum_k {\cal M}^k$) is given by 
\begin{eqnarray}\label{M2}
|\mathcal{M}|^2 &=&\frac{\left(e^4|V_{ud}| G_Ff_{\pi}g_{\gamma\gamma\eta^{(\prime)}}\right)^2}{128 \pi^4} \left[p_\nu\cdot p_\tau \left( |F_0|^2-|F_1|^2 m_\pi^2 \right) \right. \frac{}{}
+2 \left. p_\nu\cdot p_\pi \left(|F_1|^2 p_\tau \cdot p_\pi + m_\tau \textrm{Re}[F_0 F_1^*]\right)\right]\,,
\end{eqnarray}
where the $F_0=\sum_{k=a,b,d} F_0^k$ and $F_1=\sum_{k=a,b,d} F_1^k$ functions are given in terms of the invariant Passarino-Veltman (PaVe) scalar functions~\cite{Passarino:1978jh}. 
The explicit expression for these form factors can be found in appendix B of \cite{arXiv_v1}. We will provide a Mathematica file with these results upon request; they are functions of two independent kinematical scalars which 
can be chosen as $s_{12}=(p_{\pi}+p_{\eta})^2=(p_{\tau}-p_{\nu})^2$, the square of the invariant-mass of the hadronic 
system, and $s_{13}=(p_{\pi}+p_{\nu})^2=(p_{\tau}-p_{\eta})^2$. The functions $F_{0,1}$ have been obtained using the Mathematica packages FeynCalc \cite{Mertig:1990an} and 
LoopTools \cite{vanOldenborgh:1989wn, Hahn:1998yk}.

The contribution of Eq.~(\ref{M2}) alone to the branching ratio of the considered decays can be calculated straightforwardly~. 
Using the notation $BR_{P^0}^{\gamma\gamma} \equiv BR(\tau^-\to P^0\pi^-\nu_{\tau})$ when $P^0$ is produced from a $2\gamma$ intermediate state ($P=\pi,\eta,\eta'$), 
we obtain
\begin{equation}\label{BRgammagamma}
 BR_{\pi^0}^{\gamma\gamma}\,=\,5.3\cdot10^{-13}\,,\quad BR_\eta^{\gamma\gamma}\,=\,5.2\cdot10^{-13}\,,\quad BR_{\eta'}^{\gamma\gamma}\,=\,0.8\cdot10^{-16}\,.
\end{equation}
While for the $\eta^{(\prime)}$ modes the ratio between the numbers in eq.~(\ref{BRgammagamma}) and the corresponding branching fractions predicted by the tree level 
(\,$(\pi^0-)\eta-\eta'$ mixing) contributions in 
Ref.~\cite{Escribano:2016ntp} ($BR^{{\rm tree}}_\eta\sim1.7\cdot10^{-5}$ and $10^{-7}\leq BR^{{\rm tree}}_{\eta'} \leq 10^{-8}$) is at the level of $10^{-8}$, it goes further down to $10^{-11}$ for the $\pi^0$ decay mode (with respect to 
its measured branching fraction $BR_{\pi^0}\sim25\%$), which validates neglecting the one-loop contribution in the $\tau^-\to\pi^-\pi^0\nu_\tau$ decays, as we anticipated.

Next we turn to the evaluation of the branching ratio including 
the sum of  
the tree level \cite{Escribano:2016ntp} and one-loop amplitudes. For this, we note the equivalence between the 
$F_{0,+}$ form factors appearing in Ref.~\cite{Escribano:2016ntp} and our $F_{0,1}$ form factors in Eq. (\ref{amplitud}). In order to avoid confusion between the $F_0$ form factors appearing 
in both, we will use an upper index $\gamma\gamma$ for the $F_{0,1}$ form factors defined in eq.~(\ref{amplitud}).
This allows to include the electromagnetic contribution into the vector and scalar form factors by shifting $F_{+,0}^{\pi\eta^{(\prime)}}(s_{12})  \to F_{+,0}^{\pi\eta^{(\prime)}}(s_{12})+F_{+,0}^{\gamma\gamma}(s_{12},s_{13})$, where: 

\begin{equation}\label{relationFFs}
 F_+^{\gamma\gamma}\,=\,-\frac{e^4f_\pi g_{\gamma\gamma\eta^{(\prime)}}F_1}{64\pi^2}\, , \ \ \ \ \ \quad 
 F_0^{\gamma\gamma}\,=\,\frac{\frac{\textstyle  f_\pi}{\textstyle 16 \sqrt{2} \pi^2}e^4 g_{\gamma\gamma\eta^{(\prime)}}
 \left[\frac{\textstyle F_0}{\textstyle m_\tau}+\frac{\textstyle F_1}{\textstyle 2}\left(1+\frac{\textstyle \Delta_{\pi\eta^{(\prime)}}}{\textstyle s_{12}}\right)\right]}{c^S_{\pi\eta}\frac{\textstyle \Delta_{K^0K^+}^{QCD}}{ \textstyle s_{12}}}\,,
\end{equation}
with $c^S_{\pi\eta}=\sqrt{\frac{2}{3}}$ $\left(c^S_{\pi\eta^\prime}=\frac{\sqrt{2}}{3}\right)$ and $\Delta_{PQ}=M_P^2-M_Q^2$. The factor $\Delta_{K^0K^+}^{QCD}$ corresponds to the (squared) mass splitting of the $K^0K^+$ mesons which is 
due to strong interactions. Precisely $\Delta_{K^0K^+}^{QCD}\equiv m_{K^0}^2-m_{K^+}^2-(m_{\pi^0}^2-m_{\pi^+}^2)$ cancels the electromagnetic (squared) mass splitting between the kaon states and corresponds to the referred $QCD$ (squared) mass 
difference between neutral and charged kaons.

We have used the fortran version of LoopTools to compute
\begin{eqnarray}
 BR^{{\rm tree}+\gamma\gamma}_{\eta}-BR^{{\rm tree}}_{\eta} \, &\in &\,\left[-5\cdot10^{-9},2\cdot10^{-9}\right]\,, \\  BR^{{\rm tree}+\gamma\gamma}_{\eta'}-BR^{{\rm tree}}_{\eta^\prime}\, &\in & \,\left[-3\cdot10^{-12},3\cdot10^{-12}\right]\,,
\end{eqnarray}
where the difference comes mainly from the interference or tree- and loop-level contributions and --as in Ref.~\cite{Escribano:2016ntp}-- the quoted errors only arise from the uncertainties in the overall normalization factor $F_+^{\pi\eta^{(\prime)}}(0)$.

As we stated previously, the two-photon mediated amplitude considered in this paper is negligibly small compared to the dominant contribution to the $\tau^- \to \pi^-\pi^0\nu_{\tau}$ branching fraction. However, it may affect 
the branching fraction of channels with $\eta^{(\prime)}$ meson 
at the $3\cdot10^{-4}$ level~\footnote{This relative contribution may become larger if the dominant tree-level contribution 
to the \decay decays, given by the scalar form factor, has a smoother energy distribution than the one shown in Ref.~\cite{Escribano:2016ntp}.}. 
The tree level contribution to the $\eta'$ decay channel has a big uncertainty \cite{Escribano:2016ntp}, $\sim90\%$, which is dominated by the error on $F_+^{\pi\eta'}(0)$. The main modification given by the loop contributions comes from 
the interference between the former and the one-loop contribution studied here. Consequently, this interference is also affected by the big uncertainty on $F_+^{\pi\eta'}(0)$.

In this paper we have considered the one-loop contribution to the $\tau^-\to P^0\pi^-\nu_\tau$ decays induced by the $P^0\gamma\gamma$ form factors; 
to the best of our knowledge this is the first study of electromagnetic contributions to these decays. The transition form factors 
(and the $\pi$ electromagnetic form factor) have been modeled to fulfill the low- and high-energy limits of QCD (within $U(3)$ flavor symmetry for the lightest vector resonance multiplet). The proper form factors 
asymptotics has naturally rendered finite the computation of the loop integrals. We have verified that the contributions we are considering are negligible for the $\pi^0\pi^-$ channel. 
In the case of final state with an $\eta$ meson, the $2\gamma$ intermediate states contribute -at most- with  
corrections at the $10^{-4}$ level and in the case of a decay with an $\eta^\prime$ their maximum relative size can vary between $3\cdot10^{-4}$ and $3\cdot10^{-5}$ depending on the value of the 
tree level branching ratio.

It is clear that 
searches at forthcoming flavor factories will not be sensitive to effects of two-photon contributions in \decay decays~. On the one hand, SCC have not been discovered yet and even 
if \decay decays are finally measured at Belle-II it will be very difficult to achieve a 
measurement with 
a few percent accuracy  
even with the complete Belle-II data sample. Moreover, current theoretical uncertainties are huge (see Ref.~\cite{Escribano:2016ntp} and references therein), 
which prevents pinpointing New Physics effects below the $10^{-6}$ level in the branching fractions~. The quoted analysis 
only includes the errors given by $F_+^{\pi\eta^{(\prime)}}(0)$, which imply a one-order of magnitude uncertainty for the $\eta'$ decays and some $5\%$ error on the $\eta$ decays. It is difficult to quantify the error on the 
branching ratio prediction for these modes induced by the uncertainty on the couplings entering the unitarized meson-meson scattering amplitudes (which affects the dominant scalar form factor contributions) but it will 
surely dominate over the previous one. 
In addition to measurements of the branching fraction, further information on the hadronic mass as well as on angular distributions will be helpful to disentangle New Physics effects~. 
Conversely, we confirm that the contributions considered in these paper can be neglected in forthcoming SCC searches.

\section*{Acknowledgements}
This work has been supported by Conacyt Projects No. FOINS-296-2016 ('Fronteras de la Ciencia') and 236394 and 250628 ('Ciencia B\'asica'). The authors have benefited from discussions with 
Gilberto Tavares and Sergi Gonz\`alez Sol\'is.

\appendix
\section{Form factors of the tree-level amplitude}\label{A1}
 In this appendix we recall the main formulas obtained in Ref.~\cite{Escribano:2016ntp} for the \decay results at tree level, cf. eqs.~(\ref{relationFFs}).

 The amplitude of the decay $\tau^{-}\to\pi^{-}\eta^{(\prime)}\nu_{\tau}$ reads
\begin{equation}
\mathcal{M}=\frac{G_{F}}{\sqrt{2}}V_{ud}\bar{u}(p_{\nu_{\tau}})\gamma_{\mu}(1-\gamma_{5})u(p_{\tau})\langle\pi^{-}\eta^{(\prime)}|\bar{d}\gamma^{\mu}u|0\rangle\,.
\end{equation}
where the hadron matrix element is
\begin{equation}
\langle \pi^{-}\eta^{(\prime)}|\bar{d}\gamma^{\mu}u|0\rangle=\left[(p_{\eta^{(\prime)}}-p_{\pi})^{\mu}+\frac{\Delta_{\pi^{-}\eta^{(\prime)}}}{s}q^{\mu}\right]c^{V}_{\pi\eta^{(\prime)}}F_{+}^{\pi\eta^{(\prime)}}(s)+
\frac{\Delta^{QCD}_{K^{0}K^{+}}}{s}q^{\mu}c^{S}_{\pi^{-}\eta^{(\prime)}}F_{0}^{\pi^{-}\eta^{(\prime)}}(s)\,
\label{vectorcurrent}
\end{equation}
and we have used $s=q^2=(p_\pi+p_{\eta^{(\prime)}})^2$, $c^{V}_{\pi\eta^{(\prime)}}=\sqrt{2}$, 
and definitions introduced after eqs.~(\ref{relationFFs}).

Thus, the differential partial decay width, as a function of the $\pi^{-}\eta^{(\prime)}$ invariant mass, is
\begin{eqnarray}
\frac{d\Gamma\left(\tau^-\to\pi^-\eta^{(\prime)}\nu_\tau\right)}{d\sqrt{s}}\,&=&\,\frac{G_F^2M_\tau^3}{24\pi^3s}S_{EW} \left|V_{ud}F_+^{\pi^-\eta^{(\prime)}}(0)\right|^2
\left(1-\frac{s}{M_\tau^2}\right)^2\nonumber\\
& &\times  \left\lbrace\left(1+\frac{2s}{M_\tau^2}\right)q_{\pi^-\eta^{(\prime)}}^3(s)|\widetilde{F}_+^{\pi^-\eta^{(\prime)}}(s)|^2+\frac{3\Delta_{\pi^-\eta^{(\prime)}}^2}{4s}q_{\pi^-\eta^{(\prime)}}(s)|\widetilde{F}_0^{\pi^-\eta^{(\prime)}}(s)|^2\right\rbrace,
\label{width}
\end{eqnarray}
where
\begin{equation}
\widetilde{F}_{+,0}^{\pi^-\eta^{(\prime)}}(s)=\frac{F_{+,0}^{\pi^-\eta^{(\prime)}}(s)}{F_{+,0}^{\pi^-\eta^{(\prime)}}(0)},
\end{equation}
are the two form factors normalized to unity at the origin. In eq.~(\ref{width}) we have introduced the short-distance electroweak correction factor $S_{EW}=1.0201$~\cite{Erler:2002mv} and 
\begin{equation}\label{q}
q_{PQ}(s)=\frac{\sqrt{s^2-2s\Sigma_{PQ}+\Delta_{PQ}^2}}{2\sqrt{s}}\,,\quad \Sigma_{PQ}=m_P^2+m_Q^2\,.
\end{equation}

\section{Meson form factors and approximations in the computation of the loop integrals}\label{A2}
Expressions for the $\gamma\gamma\eta^{(\prime)}$ and charged pion electromagnetic form factors are required in the evaluation of the Feynman diagrams in Fig.~\ref{dia}. Noting that eq.~(\ref{TFF})
remains valid when including structure-dependent contributions if $U(3)$ flavor symmetry is assumed for the lightest resonance multiplet, the $\pi^0$ transition form factor encodes 
-under the discussed approximations- all dynamics needed to obtain the $\gamma\gamma\eta^{(\prime)}$ form factor \cite{Roig:2014uja, Czyz:2012nq}. Therefore, we will only discuss the pion form factors in the 
following.

In these limits, the structure of these form factors is \cite{RChT, Roig:2014uja}

\begin{eqnarray}\label{FF}
 \frac{F^\pi(s)}{F^\pi(0)}\,&=&\,\frac{M_V^2}{M_V^2-s}\,, \nonumber \\
 \frac{F^{\pi\gamma\gamma}(p^2,q^2)}{F^{\pi\gamma\gamma}(0,0)}\,&=&\,\frac{1}{2}\left[2+\frac{p^2}{M_V^2-p^2}+
 \frac{q^2}{M_V^2-q^2} + \frac{(p^2+q^2)M_V^2}{(M_V^2-p^2)(M_V^2-q^2)}\right]\,,
\end{eqnarray}
where $M_V\sim M_\rho(770)$ is the $U(3)$ average mass of the lowest-lying vector resonance nonet, $F^\pi (0)=1$ with an excellent approximation and 
$F^{\pi\gamma\gamma}(0,0)=g_{\gamma\gamma\pi^0}$.

We point out that the purpose of including these form factors in the evaluation of the one-loop diagrams in Fig.~\ref{dia} is two-folded: on the one hand, they incorporate 
the finite size of the pions and their structure; on the other, they naturally regulate the ultraviolet divergences appearing in diagrams $(b)$ and $(d)$ through their Brodsky-Lepage behaviour~\cite{BL}.

In our computations, we have included $F^\pi(s)$ as given in eq.~(\ref{FF}). We have noted that ultraviolet divergences always arise when both $p^2$ and $q^2$ in 
$F^{\pi\gamma\gamma}(p^2,q^2)$ tend to $\infty$. This implies that neglecting the last term of $F^{\pi\gamma\gamma}(p^2,q^2)$ does not spoil the finiteness of the result. 
Since LoopTools could not handle the case when the term with the double propagator is included, we decided  to neglect it. We justify this approximation in the following. 

We have verified that the bulk of the contribution to the loop integrals comes from the region with low photon virtualities (we understand this because the kernel of the 
integration having photon propagators). This, by the way, makes negligible the corrections induced by the finite width of the $\rho$ meson and justifies neglecting the 
contributions from excited resonance multiplets. Taking all this into account, we 
considered the following simplified expression for the pion transition form factor (which warrants the cancellation of ultraviolet divergences in our loop integrations)
\begin{equation}\label{SimplifiedTFF}
 \frac{F^{\pi\gamma\gamma}(p^2,q^2)}{F^{\pi\gamma\gamma}(0,0)}\,=\,\frac{1}{2}\left[2+\frac{a p^2 + b}{M_V^2-p^2}+\frac{a q^2 + b}{M_V^2-q^2}\right]\,,
\end{equation}
where $b=0$ (in agreement with the ABJ prediction); $a=1$ would thus correspond to neglecting the contribution of the last term of this form factor in eq.~(\ref{FF}), which does not modify sizeably 
its value in the dominant integration regions. Given the above discussion, we judge eq.~(\ref{SimplifiedTFF}) a sufficient approximation for our computations and we will use 
$a=1$ and $b=0$ in the numerics. Nevertheless, our results in appendix B of Ref.~\cite{arXiv_v1} are given in terms of $a$ and $b$. Varying $a$ will modify the coefficient of the $1/Q^2$ 
($Q=p,q$) asymptotic damping of the form factor. 
A tiny value of $b\neq0$ would still be consistent with the very small error of the $\pi^0\to\gamma\gamma$ decay rate, which is in agreement with the ABJ prediction.

Finally, we remark that diagram $a)$ is finite even using $F^{\pi\gamma\gamma}(0,0)=g_{\gamma\gamma\pi^0}$, i. e. neglecting model-dependent contributions in the transition 
form factor. Being its topology more complicated than the one in diagrams $(b)$ and $(d)$ we have followed this procedure so as to be able to evaluate it with LoopTools. Adding 
structure-dependent terms to this point-like interaction will only reduce the strength of the coupling (and thus the contribution coming from this diagram) for larger photon 
virtualities. We also note that the contribution to the branching fraction of diagram $(a)$ is subdominant with respect to that of diagrams $b)$ and $d)$. This contribution of diagram 
$a)$ alone is two orders of magnitude smaller than the figures in eq.~(\ref{BRgammagamma}). Because of this, we disregard the error induced by considering $F^{\pi\gamma\gamma}(0,0)=g_{\gamma\gamma\pi^0}$ in 
the evaluation of diagram $a)$. Incidentally, diagrams $b)$ and $d)$ give very similar contributions to the branching ratio for the $\pi^0$ and $\eta$ channels. For the $\eta'$ channel, the contribution 
of diagram $d)$ approximates the total branching ratio within 10$\%$.


\end{document}